# Correlations of atmospheric water ice and dust in the Martian Polar regions

Adrian J. Brown[*,1]

Michael J. Wolff[2]

Jeffrey D. Scargle[3]

[1] SETI Institute, 189 Bernardo Ave, Mountain View, CA 94043, USA

[2]Space Science Institute, 18970 Cavendish Rd, Brookfield, WI, 53045, USA

[3]Space Science Division, NASA Ames Research Center, Moffett Field, 94035, USA

Short running title: "Correlations of Mars polar ice and dust"

[*] Corresponding author address: Adrian J. Brown, SETI Institute, 189 N. Bernardo Ave, Mountain View, CA 94043; Tel: +1 (650) 810-0223; email: abrown@seti.org

# ABSTRACT

We report on the interannual variability of the atmospheric ice/dust cycle in the Martian polar regions for Mars Years 28-30. We used CRISM emission phase function measurements to derive atmospheric dust optical depths and data from the MARCI instrument to derive atmospheric water ice optical depths.

We have used autocorrelation and cross correlation functions in order to quantify the degree to which dust and ice are correlated throughout both polar regions during Mars Years 28-29.

We find that in the south polar region, dust has the tendency to "self clear", demonstrated by negative autocorrelation around the central peak. This does not occur in the north polar region.

In the south polar region, dust and ice are temporally and spatially anti correlated. In the north polar region, this relationship is reversed, however temporal correlation of northern dust and ice clouds is weak – 6 times weaker than the anticorrelation in the south polar region.

Our latitudinal autocorrelation functions allow us to put average spatial sizes of event cores and halos. Dust events in the south are largest, affecting almost the entire pole, whereas dust storms are smaller in the north. Ice clouds in north are similar in latitudinal extent to those in the south (both have halos < 10°).

Using cross-correlation functions of water ice and dust, we find that dust events temporally lag ice events by 35-80 degrees of solar longitude in the north and south poles, which is likely due to seasonality of dust and ice events.

Summary. We report on the interannual variability of the atmospheric ice/dust cycle in the Martian polar regions for Mars Years 28-30. We used CRISM emission phase function measurements to derive atmospheric dust optical depths and data from the MARCI instrument to derive atmospheric water ice optical depths. Using auto and cross correlation functions, we quantify the degree to which dust and ice events are correlated in the north and south polar regions of Mars.

# KEY POINTS

Key Point 1. South polar dust and ice are strongly temporally and spatially anti-correlated.
Key Point 2. North polar dust and ice are weakly temporally correlated – 6x weaker than south.
Key Point 3. We quantify southern dust "self-clearing". No north self-clearing is observed.

# 1. Introduction

We use the derived water ice and dust optical depth to explore the seasonal atmospheric conditions in the Martian polar regions and discuss several important polar science questions.

We have used Compact Reconnaissance Imaging Spectrometer for Mars (CRISM) emission phase function (EPF) measurements in the Martian poles (regions poleward of 55 deg. latitude) to map atmospheric dust optical depth as a function of season for three Mars years. We have derived the water ice atmospheric opacity using data from the MARs Color Imager (MARCI) instrument.

Using these datasets, we aim to look for relationships between two key atmospheric constituents of the polar atmosphere in a quantitative fashion by using a correlation function approach. We first give a background into past studies of the Martian polar atmospheric processes. For brevity we must omit many important atmospheric studies that did not address the polar regions.

## 1.1 Timing Conventions

In this study, we refer to timings using the abbreviation "MY" for Mars Year, specified as starting on 11 April 1955 by Clancy et al. (2000). MY 28 began 22 Jan 2006, MY 29 began 10 December 2007. The term "$L_s$", or solar longitude, refers to the celestial location of the sun when viewed from Mars (0-360). This is a convenient measure of the Martian seasons – 180-360 is southern spring and summer. Aphelion occurs at $L_s$=71 and perihelion occurs at $L_s$=251.

## 1.2 Previous Polar Atmospheric Studies

Planetwide surveys of atmospheric opacity have often omitted the areas poleward of 65° due to the variability of the polar opacity and the presence of optically thick polar hood clouds during winter (Martin, 1986; Smith, 2008). We briefly outline below the results of studies of the Martian atmosphere that have examined clouds and dust in the polar regions.

### 1.2.1 Pre-Mariner studies

Prior to human-built spacecraft arriving at Mars, the winter atmosphere in the polar regions was known to be populated with clouds, postulated to be water ice (Martin et al., 1992). In the years between Mariner 4's first successful flyby of Mars on 14-15 July 1965 and the Mariner 6-7 flybys in the first week of August 1969, work began on the first global atmospheric models of Mars (Leighton and Murray, 1966; Gierasch and Goody, 1968; Leovy and Mintz, 1969) which predicted the presence of $CO_2$ ice caps and a seasonal $CO_2$ condensation-sublimation cycle.

**1.2.2 Mariner 9 and Viking**

Mariner 9 successfully achieved Mars orbit insertion in May 1971. Onboard Mariner 9 a (thermal) InfraRed Interferometer Spectrometer (IRIS) was able to measure surface temperature, and identified $H_2O$ in the Martian atmosphere (Hanel, 1972). Five years after Mariner 9, instruments on Viking Orbiters I and II mapped the Martian globe. The Mariner 9 and Viking images were used to create a global data set of Martian cloud types (Kahn, 1984). The InfraRed Thermal Mapper (IRTM) instruments on were used to observe atmospheric temperatures (Martin, 1986; Martin and Richardson, 1993) and the Mars Atmospheric Water Detector (MAWD) was used to seasonally map $H_2O$ vapor (Jakosky and Farmer, 1982; Fedorova et al., 2004). Clancy and Lee (1991) used Viking IRTM Emission Phase Functions to derive ice cloud and dust albedos in several locations around the planet and propose a dust single scattering albedo of 0.92 in the visible wavelength range for "global" dust. Forget et al. (1995) used Mariner 9 IRIS and Viking IRTF to infer possible $CO_2$ snowfall in the polar regions.

**1.2.3 Telescopic Studies**

Terrestrial telescopes have continued to provide important data on the atmosphere of Mars. Clancy et al. (Clancy et al., 1996) used the Very Large Array, Kitt Peak and Hubble Space Telescope (HST) data to identify the aphelion cloud belt (ACB) and determined the Martian northern atmosphere was saturated with water vapor during that period. Wolff et al. (1997) used HST data to derive atmospheric opacities during 1995 Martian dust storms. Clancy and Sandor (1998) used submillimeter measurements from the James Clerk Maxwell Telescope on Mauna Kea to observe surprisingly cold temperatures at 50-80km above the surface, from which they inferred the presence of high altitude $CO_2$ ice clouds. Glenar et al. (2003) used Kitt Peak data to observe water ice clouds contemporaneously with the spacecraft Mars Global Surveyor (MGS).

**1.2.4 Thermal Emission Spectrometer (TES) on MGS**

The Thermal Infrared Spectrometer on Mars Global Surveyor provided further atmospheric mapping of Mars from 1997 to 2006. Pearl et al. (2001) documented water ice clouds for the first year of TES operations. TES EPFs were used by Clancy et al. (2003) to derive water ice cloud opacities and particle sizes as a function of season. Wolff and Clancy (2003) used TES measurements to place constraints on the size of Martian dust particles. Smith (2004) compared interannual water vapor, dust and water ice opacities and atmospheric temperature profiles. Eluszkiewicz (2008) derived dust and water ice optical depth for 148 TES spectra at 87N during an unspecified fall and winter period and obtained dust optical depths from 0-0.25. Horne and Smith (2009) also analyzed interannual TES data over the polar regions and reported dust optical depths of 0-0.5 as typical over both poles. They reported the presence of annular bands of dust and ice opacity in the north pole, with no such patterns in the south. Horne and Smith's retrieval method left gaps in transition regions near the cap edge due to difficulties in modeling the surface temperatures of this region.

TES data were also used (in combination with Viking IRTM) to compare water vapor variability above the Martian north pole (Pankine et al., 2009, 2010). Pankine et al. found that twice as much water vapor was present above the north polar cap during the MGS TES observation period than during Viking observations, perhaps due to differences in atmospheric circulation patterns.

### 1.2.6 Mars Orbiting Camera (MOC) on MGS and Thermal Emission Imaging System (THEMIS) on Mars Odyssey

In addition to the (primarily infrared) spacecraft measurements of the Mars atmosphere, a rich history of dust storm activity have been documented using the Viking Orbiter cameras (James et al., 1979; James, 1982) and Mars Orbiter Camera (MOC) on MGS (James et al., 2001; Cantor et al., 2001, 2002; Cantor, 2007; Cantor et al., 2011).

Wang and Ingersoll (2002) used the daily global maps of Mars produced by the MOC camera to map the appearance of polar hood clouds, distribution of lee waves and streak clouds in both poles. Wang et al. (Wang et al., 2005) and Wang and Fisher (2009) mapped cyclonic activity in the north pole for MY24-28 that included fronts and spiral structures of baroclinic clouds that apparently do not occur in the south pole.

Inada et al. (2007) used the visible channel of the THEMIS instrument to monitor atmospheric activity for MY26-27. They observed katabatically driven south polar trough clouds and north polar dust plumes, and mesospheric clouds.

### 1.2.7 The Mars Orbiter Laser Altimer (MOLA) on MGS and the Phoenix Lidar

The MGS Spacecraft carried the LIDAR MOLA instrument which made extensive measurements of the Martian atmosphere, including thickness of clouds in the north polar hood and potential detections of $CO_2$ clouds in the polar regions (Pettengill and Ford, 2000; Hu et al., 2013). In addition to the orbital MOLA instrument, the Phoenix lander, which landed in the northern polar regions, carried an upward pointing lidar which has been used to learn more about clouds (Whiteway et al., 2009) and dust, particularly in the planetary boundary layer (lower 4km) of the polar regions (Komguem et al., 2013).

Recently, an improved multiwavelength, polarization sensitive LIDAR has been proposed that would answer many of the scientific questions posed in this paper (Brown et al., 2014a).

### 1.2.8 SPectroscopy for Investigation of the Characteristics of the Atmosphere of Mars (SPICAM) and Observatoire pour la Minéralogie, l'Eau, la Glace et l'Activité

**(OMEGA) on Mars Express**

Fedorova et al. (2009) used SPICAM to derive water ice and dust loadings in the northern hemisphere from MY28 $L_s$=130-160. Montmessin et al. (2006) have observed high altitude (100km) $CO_2$ ice clouds using SPICAM. Forget et al. (2009) used SPICAM to measure seasonal temperatures and density of the upper atmosphere of Mars (60-130km above the surface). Mateshvili et al. (2007, 2009) used SPICAM UV wavelengths to examine water ice clouds in the north and south polar hoods and the aphelion cloud belt. Trokhimovsky et al. (2014) presented a summary of SPICAM observations for MY27-31, showing peaks of 60-70 pr microns in the north and 20 pr microns in the south.

The OMEGA near infrared spectrometer has been used by Vincendon et al. (2008) to derive dust opacities over the south polar ice cap during spring and summer. Vincendon et al. (2009) also investigated OMEGA EPF data over dark regions of Mars and found them consistent with dust particles with sizes between 1-2 micron diameter. Maattanen et al. (2011) used OMEGA and High Resolution Stereo Camera (HRSC) data to map the occurrence and morphology of mesospheric $CO_2$ ice clouds. They infer convection may be responsible for the clumpy appearance of these clouds.

Doute et al. (2013) developed a method called *MARS-ReCO* that derives surface albedo from an EPF measurement, much like the method we outline below. The key difference is their approach uses a non-Lambertian phase function to model the reflection from the surface. They applied this method (Ceamanos et al., 2014; Fernando et al., 2014) to CRISM and OMEGA observations to remove the affects of the atmosphere – here we seek to remove the effects of the surface and study just the atmosphere.

### 1.2.9 Martian Climate Sounder (MCS) on MRO

The MCS is a multispectral, multi-camera observation platform that looks ahead of the spacecraft to observe the Martian atmosphere at discrete angles covering the lower Martian atmosphere. MCS was designed to map the seasonal changes in the Martian atmosphere (McCleese et al., 2007). The current MCS retrievals use a single scattering assumption which limits their applicability to above 10km from the Martian surface (Kleinböhl et al., 2009).

Hayne et al. (2012) used MCS data to map extensive tropospheric clouds over colds spots on the south polar winter cap, indicating the colds spots are caused by snowfall. Hayne et al. (2013) estimated 3-20% of mass deposited in the south pole is by snowfall.

A major achievement of the MCS program was the first comprehensive mapping of the north (Benson et al., 2011) and south (Benson et al., 2010) polar hoods. Benson et al. found the north polar hood water ice clouds were longer lived and thicker in the north than the south. In the north, clouds were present from $L_s$=150 (late summer) to $L_s$=30 (late spring). In the south, clouds were present in two phases, $L_s$=10-70 (fall) and $L_s$=100-200 (winter). In the north, Benson et al. found the north polar hood formed as a single cloud deck, whereas in the south, two cloud decks, separated by a temperature inversion, were present.

**1.2.10 CRISM visible and infrared imaging spectrometer on MRO**

CRISM data has also been used, primarily in the mid-latitude regions, to determine the single scattering albedo of the 'global dust' (Wolff et al., 2009). CRISM has also been recently used to derive global carbon monoxide and water vapor measurements (Smith et al., 2009; Toigo et al., 2013) and occasional vertical dust limb measurements (Smith et al., 2013). Guzewich et al. (2014) recently used CRISM limb profiles measurements to establish that atmospheric dust particle sizes have an effective radius of 1 micron for most of the atmospheric column below 40km, and that in the polar hoods, water ice particles have uniform size, with an effective radius of 1.5 microns.

CRISM has also been used to map the surface $CO_2$ and $H_2O$ ice cap springtime recessions for the north (Brown et al., 2008a, 2012) and south (Brown et al., 2010) polar caps primarily using spectral absorption band analysis techniques (Brown, 2006; Brown et al., 2008b). In addition, Brown et al. (2014b) made measurements of 1-6 microns of water ice deposition in the south polar cap each summer, which is a critical factor in the Martian global energy budget and enables improvements to models of the global Martian climate.

**1.2.11 Global Climate Model studies of polar dust and water clouds**

Burk (1976) used a parameterized model of the Martian south polar cap to investigate the ability of the on-cap winds to lift dust prior to global dust storms. Siili et al (Siili et al., 1997) conducted a similar study. Haberle et al (1979) used a 2D model of the Martian polar cap to examine the polar wind dynamics. Pollack and Haberle (1990) used a 3D GCM to investigate the effect of dust on the condensation of $CO_2$ ice in the polar regions, but did not address the effect of clouds.

Richardson and Wilson (2002a) used a WRF GCM model to simulate the Martian water cycle. In the polar regions, their study highlighted the effects of polar vortex walls preventing volatiles from moving over the cap during autumn and winter. Richardson et al. (Richardson and Wilson, 2002b) used a GCM study to highlight the role of topography in forcing the north polar region to remain relatively water rich, compared to the south polar region.

Montmessin et al. (2004) used the LMD GCM to investigate global clouds, with some emphasis on polar hood clouds. Interestingly, Figure 9 and 10 of Montmessin et al. show models of summer daytime polar hood clouds that extend off the polar caps, in the region of 50-80°N, $L_s$=60-120. Indications of these clouds have been measured by the Mars Orbiting Laser Altimeter (Neumann et al., 2003)**.**

Tyler and Barnes (2005) used a mesospheric climate model to investigate the dynamics of the water cycle in the north polar region during summer. They found that strong but transient eddies are present throughout summer, and that that these likely are important in the water cycle of the north polar cap.

## 2. Methods

### 2.1. CRISM EPF Measurements

CRISM has the ability to take 'gimballed' observations of the surface as it passes over a target, thus creating what is termed an Emission Phase Function 'EPF' measurement (Figure 1) (Murchie et al., 2007). We report here on our initial investigations of the EPF polar observations and our attempts to model dust and ices suspended in the atmosphere and soil and ice covered surface.

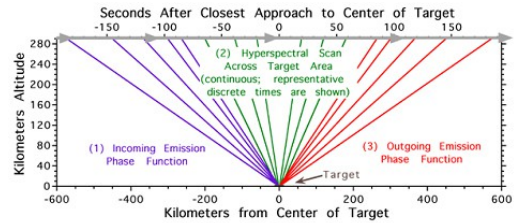

**Figure 1.** Schematic of a CRISM EPF observation (Murchie et al., 2007).

Table 1 shows all the CRISM EPF observations poleward of 55°. CRISM is limited to daytime observations and MRO is in a ~250km circular orbit that crosses the equator south to north at 1500 local Mars standard time.

### 2.2. CRISM Dust Retrieval Methods

We used the DISORT (Stamnes et al., 1988) algorithm to simulate the interaction of the Martian atmosphere and surface with the incident solar radiation with the goal of extracting three parameters: surface Lambert albedo, dust and ice aerosol optical depth. The retrieval is accomplished by fitting a CRISM EPF (in a least-squares sense) at a single wavelength (0.696 micron).

We used an elevation corrected scheme for Rayleigh molecular scattering scaling factor based on a 10km scale height (Pollack et al., 1979). We have used the MOLA heights in the center of each observation to scale the atmospheric surface pressure and assumed no barometric pressure changes with height.

Starting with a Martian atmosphere with 39 2km high layers from 1-79km (with a base layer of 1km and a top layer of 100km), with 10km scale height (Table 2) we interpolate 15 layers at 50, 40, 35, 30, 25, 20, 17, 14, 11, 9, 7, 4, 2, 1, 0.00145 and 0 km. We use MOLA heights from the CRISM DDR files to scale the atmospheric column. Dust is restricted to heights of 0-80km, water ice clouds are restricted to 25-80km. Both dust and water ice are assumed to be evenly distributed within these height restrictions. We use standard corrk files (produced by M.D. Smith) for CO2, (95.3%) molecules (we currently omit seasonally variable CO, $H_2O$ and $CH_4$) and a typical ozone column at perihelion ($L_s$=240) varying from $2.45 \times 10^{-7}$ (at 6.96mbar) to $7.8 \times 10^{-5}$% at 0.02mbar. We do not treat thermal radiation from the surface or within the atmosphere (this has minimal effects at ~0.7 m).

Computational efficiency is obtained by using the a "look-up table" approach, where we have pre-computed (millions of) radiative transfer models that span the possible range of optical depth and aerosol values.

**2.2.1 Dust Error Analysis**

In general, using the diagonal element of the covariance matrix provides the most realistic estimate of the error associated with an individual CRISM dust optical depth value. However, comparisons to CRISM EPFs near the MER rovers highlights the need to consider to limiting cases (Wolff et al., 2009). Ultimately, one can calculate the error by selecting the largest value from the following list: a.) formal retrieval uncertainty (covariance matrix), b.) 10% of the optical depth, c.) 0.05.

| Earth DOY | MY/L$_s$ | EPF | FRT | HRL | HRS |
|---|---|---|---|---|---|
| 06_272-298 | 28/[113-125) | *21* | 6 | 5 | 1 |
| 298-312 | [125-132) | | | | |
| 312-326 | [132-139) | *10* | 8 | 2 | 13 |
| 326-340 | [139-146) | 6 | 7 | 1 | 1 |
| 340-354 | [146-153) | 15 | *13* | 4 | 7 |
| 354-07_003 | [153-161) | 20 | 3 | 2 | |
| 003-017 | [161-168) | 24 | 6 | 2 | 1 |
| 017-031 | [168-176) | 2 | 1 | 3 | 1 |
| 031-048 | [176-185) | 11 | 3 | 1 | |
| Northern winter | | | | | |
| --- | --- | 36 | 1 | 2 | 3 |
| 255-269 | [312-320) | *14* | 7 | 1 | |
| 269-283 | [320-328) | | 2 | 4 | 1 |
| 283-297 | [328-335) | | 1 | 2 | |
| 297-311 | [335-344) | | 1 | 1 | |
| 311-325 | [344-351) | | | | |
| 325-339 | [351-358) | | | | |
| 339-353 | [358-005) | | 9 | | |
| 353-08_002 | 29/[5-12) | | 38 | | |
| 08_002-016 | [12-19) | | 75 | 9 | 1 |
| 016-030 | [19-25) | | 122 | 10 | 3 |
| 030-044 | [25-32) | | 144 | 8 | |
| 044-058 | [32-38) | | 120 | 18 | 7 |
| 058-072 | [38-44) | | 159 | 18 | 38 |
| 072-086 | [44-50) | | 9 | 1 | |
| 086-100 | [50-56) | | | | |
| 100-114 | [56-62) | | 50 | 4 | 7 |
| 114-128 | [62-69) | | 74 | 38 | 16 |
| 128-142 | [69-75) | | 29 | *15* | 2 |
| 142-156 | [75-81) | | 13 | 1 | |
| 156-170 | [81-87) | | 44 | 12 | |
| 170-184 | [87-93) | | 53 | 17 | 2 |
| 184-198 | [93-100) | | 43 | 59 | 12 |
| 198-212 | [100-106) | | 59 | 19 | 4 |
| 212-226 | [106-112) | | 22 | *34* | 9 |
| 226-240 | [112-119) | | 46 | 25 | 25 |
| 240-254 | [119-125) | | 47 | 14 | 21 |
| 254-268 | [125-132) | | 22 | 12 | 22 |
| 268-282 | [132-139) | | *32* | 4 | 7 |
| 282-296 | [139-146) | | 32 | 24 | 14 |
| 296-310 | [146-153) | | 21 | 22 | 11 |
| 310-325 | [153-160) | | 39 | 4 | 17 |
| 356-09_012 | [177-190) | | 4 | | |
| Northern winter | | | | | |
| 193-233 | [301-325) | | 2 | 1 | 1 |
| MRO in safe mode | | | | | |
| 10_031-039 | 30/[45-49) | | 37 | 15 | |

| Earth DOY | MY/L$_s$ | EPF | FRT | HRL | HRS |
|---|---|---|---|---|---|
| 06_352-5 | 28/[152-159) | 8 | | | |
| 006-016 | [160-168) | 9 | | | |
| 016-033 | [168-176) | 3 | | 3 | 1 |
| 033-044 | [176-183) | 10 | 6 | 3 | 4 |
| 044-059 | [183-192) | 20 | 17 | 1 | 6 |
| 059-073 | [192-200) | 8 | 20 | 1 | 5 |
| 073-086 | [200-208) | 2 | 14 | | 1 |
| 086-101 | [208-217) | 25 | 27 | | 24 |
| 101-115 | [217-225) | *17* | 29 | 11 | 12 |
| 115-116 | [225-234) | | 4 | 1 | |
| 132-142 | [234-243) | | 36 | | 3 |
| 142-156 | [243-252) | 27 | 42 | 5 | 8 |
| 156-171 | [252-261) | 3 | 37 | 7 | 2 |
| 171-185 | [261-270) | *12* | 48 | 4 | 1 |
| 185-198 | [270-278) | 124 | 30 | 2 | 2 |
| 199-212 | [278-286) | *43* | 30 | | |
| 213-225 | [286-295) | *49* | 137 | 1 | 6 |
| 227-240 | [295-303) | 22 | *111* | 2 | 2 |
| 241-255 | [303-312) | | *111* | 4 | |
| 255-269 | [312-320) | *16* | 66 | 3 | 4 |
| 269-283 | [320-328) | | 34 | 16 | 4 |
| 283-297 | [328-335) | | 49 | 5 | 13 |
| 297-311 | [335-344) | | 18 | 8 | 4 |
| 311-348 | [344-002) | | | | |
| 348-00 | 29/[002-012) | | 61 | 1 | 2 |
| 08_004-33 | [012-026) | | 43 | 3 | |
| 033-074 | [026-044) | | 39 | 3 | 2 |
| Southern winter | | | | | |
| 356-09_001 | [177-184) | | 8 | 2 | |
| 019-033 | [194-202) | | 14 | 1 | |
| 034-050 | [203-211) | | 13 | 8 | 16 |
| 051-067 | [214-223) | | 8 | 4 | 2 |
| 069-082 | [223-233) | | 46 | 17 | 10 |
| 083-097 | [233-242) | | 31 | 18 | 10 |
| 098-112 | [242-252) | | 28 | 30 | 4 |
| 113-126 | [252-261) | | 16 | 14 | 9 |
| 128-142 | [261-271) | | 37 | 8 | 10 |
| 143-154 | [271-278) | | 14 | 16 | 5 |
| 162-175 | [282-292) | | 15 | 2 | |
| 176-192 | [292-302) | | 16 | 14 | 12 |
| 193-206 | [302-310) | | 28 | 14 | 17 |
| 207-217 | [310-317) | | 15 | 20 | 5 |
| 223-237 | [319-327) | | 41 | | |
| Southern winter followed by MRO in safe mode | | | | | |
| TOTAL | n=2177 | 399 | 1319 | 252 | 207 |

**Table 1.** Totals of CRISM observations relevant to this study. North polar observations are on the left, and south polar on the right. Counts in italics indicate some missing geometries. Each line corresponds to the two week MRO planning cycle. DOY column gaps are when CRISM collected no data at the south pole.

### 2.3 MARCI Ice Opacity Retrieval methods

The Mars Color Imager (MARCI) instrument onboard the Mars Reconnaissance Orbiter (MRO) obtains near-global coverage of Mars on a daily basis (Malin et al., 2007). Of specific interest for studies of water ice clouds is the presence an ultraviolet (UV) band centered near 320 nm that exploits the reduced surface contrast and increased contribution scattering by ice aerosols (particularly compared to the much darker dust aerosols). These aspects of the atmosphere, where combined with the well calibrated nature of the MARCI camera, even in the UV (Bell et al., 2009; Wolff et al., 2010), allow for the robust retrieval column-integrated ice optical depths on a daily (and global) basis. While a manuscript providing the proverbial "gory details" of the retrieval and its uncertainties is presently being written, a reasonable description of the algorithm and its components have already appeared in several publications (Wolff et al., 2011, 2014). In addition, a public release of the ice cloud maps has been developed (Wolff, n.d.; Wolff et al., 2014, 2013) of the retrieval has been presented previously. The current version of the retrieval (Wolff et al., 2014) has uncertainties in the derived water ice column in the range 0.02-0.03 with the detection threshold of about 0.03.

## 2.4 Model Atmospheric Properties

We use phase functions for water ice 'Type 1' (non-aphelion ice cloud) with 64 moments (Clancy et al., 2003) and dust particle phase function with 64 moments derived in (Tomasko et al., 1999). The dust and ice particles are assumed spherical, with a gamma particle size distribution.

For dust: $R_{eff}$=1.5 microns, $v_{eff}$=0.4, ice: $R_{eff}$=2.0 microns, $v_{eff}$=0.1. Dust optical constants were from Wolff et al. (2009) and water ice optical constants are obtained from Warren (1984). The single scattering albedo for each atmospheric layer can then be derived.

## 2.5 Correlations of Dust and Ice with discrete correlation function

In this section we examine the spatial and temporal relationships between the CRISM dust and MARCI ice opacities in the polar regions. Based on the procedure for discrete correlation functions described by Edleson and Krolik (1988) we have computed estimates of the auto correlation function within and cross correlation between these two data sets. Specifically we carried out this procedure for various subsets of the data based on the latitude and longitude of the position on the planet to which the measurements refer.

Edelson and Krolik (1988) developed what has become known as the discrete correlation function to study correlations in datasets consisting of measurements at arbitrary times, as opposed to the evenly spaced times needed for the classical correlation functions. It is easy to extend this method to estimate cross-correlations, and of course the correlation can refer to temporal or spatial variables - or to any other variance parameter (i.e. the independent variable in the functions to be correlated) for that matter.

For the same reason that applies to the case of evenly spaced data the Edelson and Krolik autocorrelations are independent of the direction of the "arrow of time" - whether time runs forwards or backwards - and symmetric around zero lag.
This procedure has been recently applied to several astronomical time series projects (Scargle, 2010; Scargle et al., 2013). Of course the cross-correlation is in general not symmetric, and provides information about to what extent one of the analyzed series leads or lags the other.

When carrying out the correlation analysis, we filtered the CRISM dust and MARCI ice datasets firstly by only including MARCI data where CRISM data was available. Then we discarded CRISM data before 1 Jul 2010 due to problems with the CRISM rotation gimbal that required half of the EPF sequence to be removed. This led to a large spread in dust retrievals after this date (compare MY30 to the previous two MY in Figure 2a and b).

Only data from the polar regions (defined as all regions poleward of 55°N and 55°S) was used in the correlation analysis. In addition, due to difficulties retrieving MARCI ice opacity over icy regions, the MARCI ice datasets were not used poleward of 80°N and poleward of 75°S. Unsuccessful MARCI water ice retrievals were also filtered out, these were typically recognized with a value of zero.

## 3. Observations

The $\tau_d$ results for CRISM EPF and the corresponding MARCI $\tau_i$ retrievals from Mars Year 28/29/30 (2006-2011) are shown in Figure 2a and 2b.

### 3.1. Polar Atmospheric Dust and ice opacity

**Interannual Repeatability**. The CRISM EPF dataset south polar dust activity is far more pronounced and repeatable than the north polar data. The south pole dust activity is more sharply concentrated each year around $L_s$=270 (southern vernal equinox). The north pole has a similar peak at $L_s$=90 (northern vernal equinox). MY28 was considerably more dusty than MY29 and 30 in both poles.

**Flushing of southern dust storms.** The southern polar dust activity shows a unique tendency to build up to a big storm period, and then decrease in activity. This is thought to be related to the dust recharge time in the southern polar region.

**Background dust levels**. The results shown in Figure 2 are consistent with background dust opacities of $\tau_d$=0.3-0.5 for the south polar region, with average excursions to 1.4 during the MY28 dust event. In the north polar region, background dust opacities also hover around $\tau_d$=0.3. This is observed as a "gap" between 0-0.3 in Figure 2a (top), particularly apparent during the summer ($L_s$=90-180) period. There are some occasional observations of even lower dust opacity displayed, although they are less common.

**Maximal Dust and Ice opacity levels and periods**. From Figure 2, we can make the following statements about peaks in ice and dust activity. Full column dust optical depths peak at $\tau_d$=4 in south at $L_s$=270 and full column dust optical depths peak at $\tau_d$=4.5 in north at $L_s$=90. Full column ice optical depths peak at $\tau_i$=2.0 in south at $L_s$=210 and full column ice optical depths peak at $\tau_i$=1.5 in north at $L_s$=45.

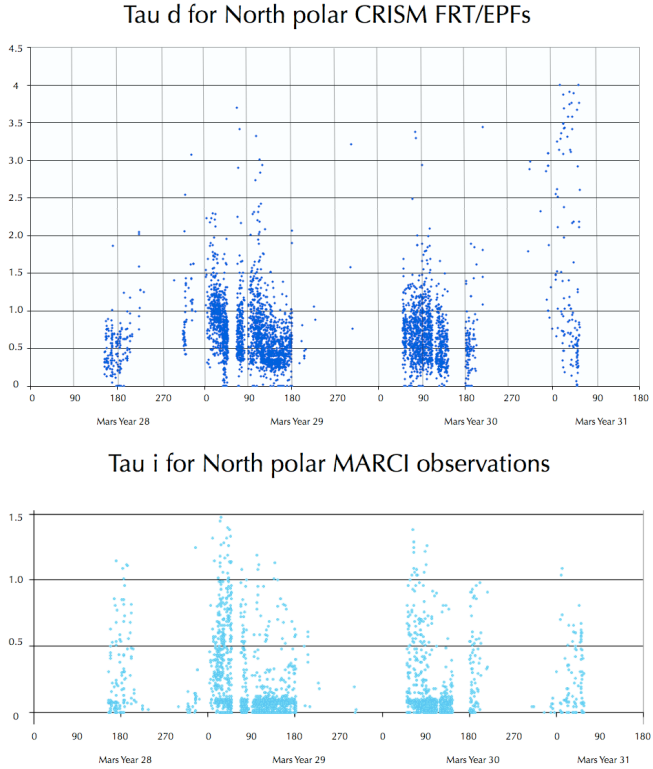
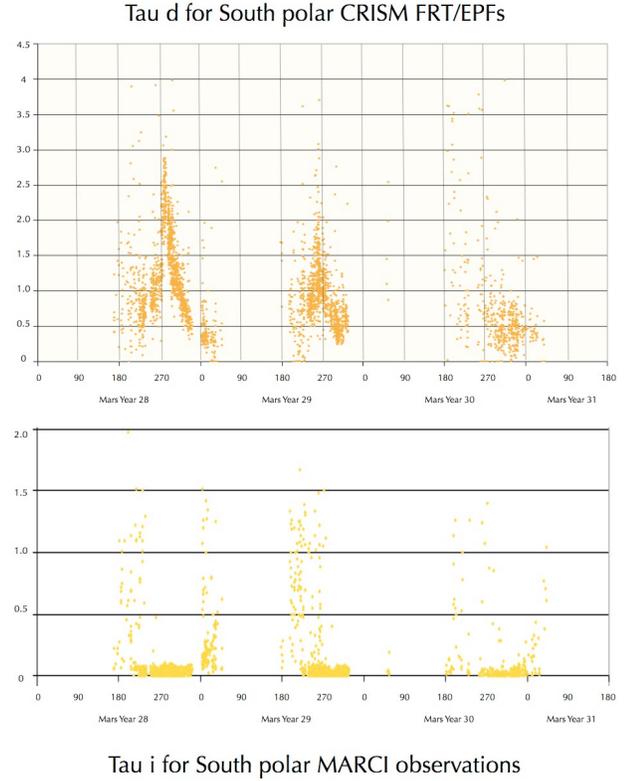

**Figure 2a.** North polar $\tau_d$ and $\tau_i$ estimates for CRISM EPF/MARCI observations (poleward of 55°N) for MY28-30.

**Figure 2b.** South polar $\tau_d$ and $\tau_i$ estimates for CRISM EPF/MARCI observations (poleward of 55°S) for MY28-30.

### 3.2 Correlations of dust and ice in the polar regions

As discussed in the methods section, we used the discrete correlation function of Edelson and Krolik (1988) to examine the auto and cross-correlations of the CRISM dust and MARCI ice opacities with time, latitude and longitude - that is, with these three quantities as independent variables and the opacities as dependent variables. There are therefore 12 auto-correlations: dust and ice (x2) and north and south poles treated separately (x2) making up 4 panels in three figures with time, longitude and latitude as independent variables (3x); and 6 cross correlations of dust vs. ice: north and south poles treated separately (x2) with time, longitude and latitude as independent variables (3x).

In computing all the correlation functions shown here the mean values of the data series have first been subtracted. We have plotted a range of the independent variable large enough to cover the significant behavior, with the plotted function typically decaying to zero correlation at large lags. The values of the ACFs are mean values for the square of the optical depth (dimensionless); these values have not been renormalized and can therefore be directly compared across all of the presented cases. In terminology standard in some areas we are presenting auto- and cross- covariance functions. However we here use the term "correlation function" here for values of:

$$C(\tau) = X(t)\,Y(t+\tau) \tag{1}$$

averaged over *t,* the cross correlation of *X* and *Y* (auto-correlation if *X=Y*).

For the ACFs we show both positive and negative lags, but of course these functions are totally symmetric and convey no information about the sense of time (time reversing the input data leaves the ACF invariant). Since 1-sigma errors are quite small on the scale of these figures, the plotted error bars are ±3 sigma, (determined as the ensemble standard deviation of the values contributing to the ACF estimate at the given lag) and are not based on the reported observational errors.

Assuming the observational errors are independent of each other, their contribution to the ACF at all nonzero lags averages to zero. At zero lag the computed ACF is the sum of two (estimated) variances: that of the underlying time series and that of the observational errors. In all cases the former dominates the latter, indicating that the optical depths are truly random processes with a well determined variance.

### 3.2.1 Autocorrelation functions (ACFs)

**Temporal autocorrelations**. The results of the temporal autocorrelation of the CRISM dust and MARCI ice datasets for north and south poles are shown in Figure 3a. All have a well defined maximum peak at zero lag, as discussed above. In addition there are varying degrees of positive correlation at small lags (<= approximately 10° of solar longitude) and in some cases positive and negative correlations at larger lags. The statistical significance of these excursions can be judged by noting the sizes of the indicated error bars. Such excursions are indicative of a "memory" in these time dependent processes, due to values at one time being correlated with values at other times.

In particular the negative values of the temporal ACF for south pole dust over the range of lags between 70 and 110 degrees of solar longitude (0.2-0.3 Earth years) would suggest a trend for dust to clear approximately a quarter of a year after it appears. The reader should keep in mind that, since the data are here effectively averaged over latitude and longitude, this trend should be interpreted in the sense of a spatial average.

**Latitude autocorrelations**. The latitude autocorrelations are presented in Figure 3b. In all cases most of the correlation is confined to within 1 or 2° of zero lag, indicating the corresponding dust and ice storms have cores of roughly this size. The ice storms on the other hand have lower lever extensions to ~10° from their centers.

Dust storms are more dispersed than ice clouds in both poles, with ~10° halos surrounding a central core. This is typical for baroclinic waves observed in the Martian north pole (Wang and Fisher, 2009). Southern dust events have a wider extent than their northern cousins, and are typically more intense (Cantor et al., 2011; Montabone et al., 2014). This is exemplified by the intense southern dust events in MY 28 and 29 which are plainly visible in Figure 2b. This is discussed further in Section 4 below.

Again, these conclusions must be interpreted in a time-averaged and longitude-averaged sense.

**Longitudinal autocorrelations**. The longitudinal autocorelations are shown in Figure 3c. The autocorrelations in all four cases are sharp peaks with little or no extension, indicating that the longitudinal spread of both dust and ice events at both poles is no more than 1-2 degrees. Note that the amplitude of the peak dust correlation at zero lag is 4 or 5 times larger (in units of optical depth squared) than that of ice, indicating that the optical depths are a factor of 2 or so larger. These conclusions for the longitudinal autocorrelation functions must be interpreted in a time-averaged and latitude-averaged sense.

The longitude ACFs are much narrower than the latitude ACFs simply because the variations over longitude (0-360) in the dataset are far larger than the latitude variations (55-90).

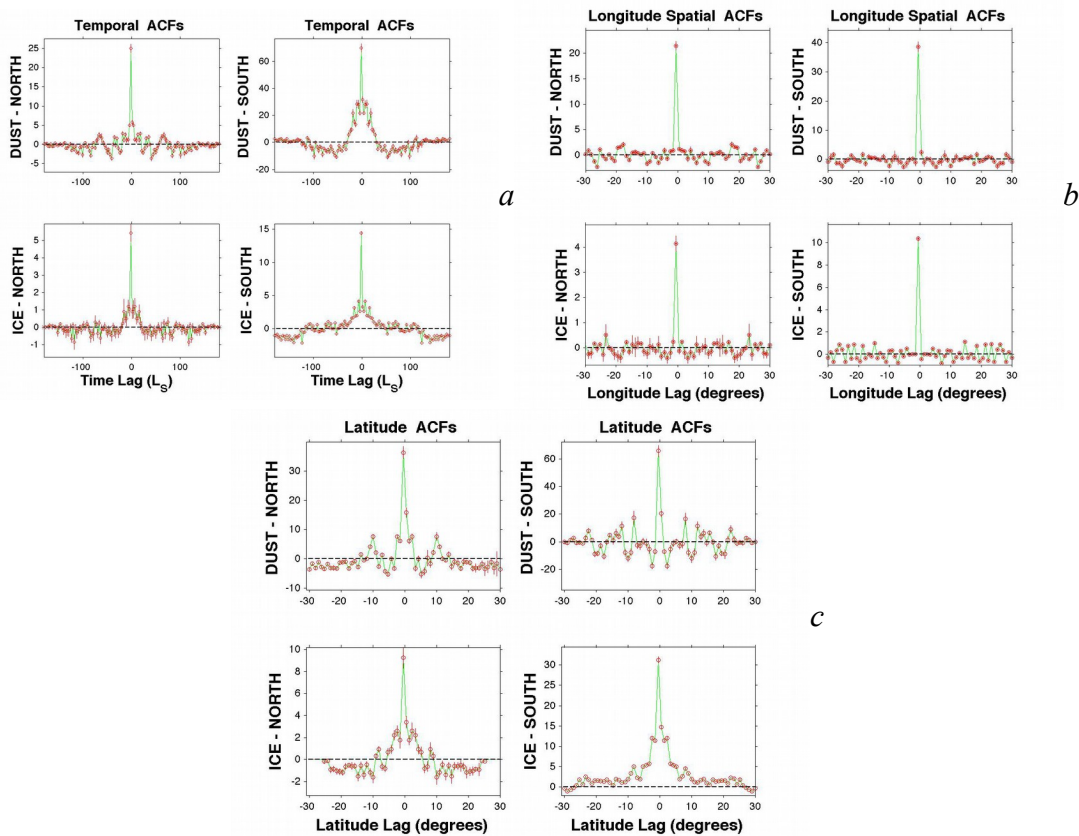

Figure caption 3a: four temporal autocorrelation functions for the dust optical depth (top row) and ice (bottom row), and the region centered on the north (left column) and south (right column). The lags refer to time as the independent variable, but expressed in terms of the difference in solar longitude at the times of the pairs of observations contributing to the ACF estimate. The correlations were computed in 128 bins covering the range from -180 to +180 degrees solar longitude, or a total of one Earth year.

Figure caption 3b: autocorrelation functions for the dust and ice optical depths near the north and south poles, laid out just as in Figure 3a, but with Martian latitude as the independent variable. The correlation function is computed in 64 bins covering the range from -30 to +30 degrees of latitude difference.

Figure caption: autocorrelation functions for the dust and ice optical depths near the north and south poles, laid out just as in Figures 3a and 3b, but now with Martian longitude as the independent variable. The correlation function is computed in 64 bins covering the range from -30 to +30 degrees of longitude difference.

### 3.2.1 Cross-correlation functions (CCFs)

**Temporal cross-correlations**. Figure 4a shows the dust and ice temporal cross correlations in the north and south poles. The weak positive peak in the north and stronger negative peak in the south, both near zero lag, indicate simultaneous dust-ice correlation and anti-correlation, respectively. Essentially, this result tells us that dust and ice are more compatible in the north and even more strongly incompatible in the south, by a factor of 6.

The physical reason for this is the amount of water ice in each hemisphere. The south is starved of water ice and the ice clouds are thin. Dust storms are also more intense in the south. Put together, this explains the temporal cross-correlation result.

More challenging to explain is that at both poles there is also indication of positive correlation in which the dust variation lags behind that of the ice by 35 to 80 degrees of solar longitude (approximately 0.1 to 0.2 Earth years). This is likely just the signature of seasonal variability in dust storms and cloud events. The peaks a sharper in the south, indicating more tightly clustered events.

**Latitude cross-correlations**. Figure 4b presents the dust and ice latitudinal cross correlations in the north and south poles. This figure indicates that in the north the dust lags the ice in latitude (as well as in the temporal sense, as discussed above). Physically, this corresponds to dust events closer to the pole by about 8° of latitude relative to ice events, and is likely related to flushing dust events coming off the polar cap during late summer when early clouds of the north polar hood are present (Calvin et al., 2014).

The dust and ice cloud events in the south appear to be uncorrelated in latitude, except for a marked anti-correlation at zero lag. That is to say the presence of dust is correlated with absence of ice at the same latitude, and vice versa. This same relation may hold in the north, since there is a weak negative peak at zero lag in the top panel. Again these statements refer to time- and longitude-averaged values.

**Longitudinal cross-correlations**. Figure 4c shows the dust and ice longitudinal cross correlations in the north and south poles. These demonstrate that the longitudes of dust events and ice events are not correlated, except for anti-correlation between dust and ice events at the same longitude in the south polar regions. This was also found for the latitude correlations in Figure 4b. in the south. In other words the presence of dust is correlated with absence of ice at the same longitude and latitude in the south and only weakly in latitude and not at all in longitude in the north. These statements refer to time- and latitude-averaged values.

The physical interpretation of the spatial anti-correlation in the north (weak) and the south (strong) is likely related to the fact that intense dust storms typically come at energetic times in the Martian seasonal cycle, and ice clouds typically enshroud the poles during energy poor winter periods.

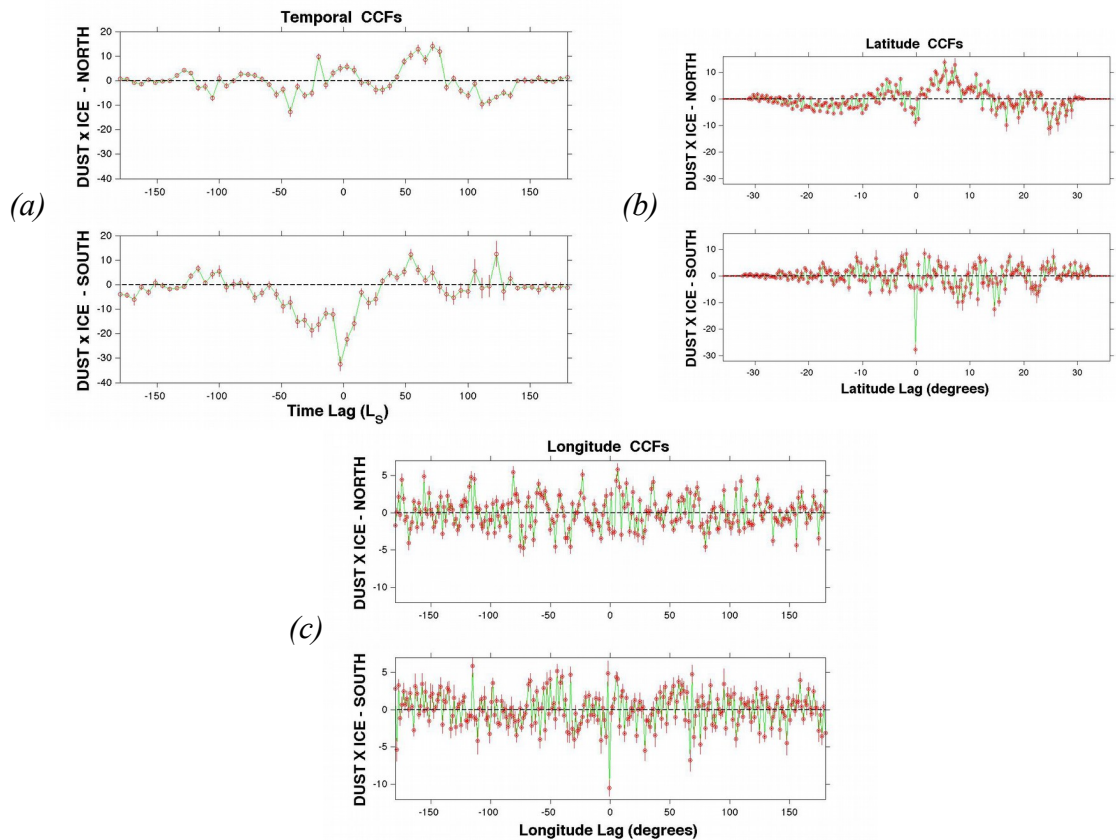

**Figure 4.** Figure caption 4a: Temporal cross correlation of dust and ice optical depth values in the north (top) and south (bottom) regions (defined as in Figs. 3a,b,c). A positive lag means that the dust variation lags behind (occurs at later times) than the ice variation. The correlations were computed in 64 bins covering the range from -180 to +180 degrees solar longitude, or a total of one Earth year. The scale of CCF amplitudes (optical depth squared) are the same in both panels, so the degrees of correlation can be directly compared.

Figure caption 4b: Latitude cross correlation of dust and ice optical depth values in the north (top) and south (bottom) regions (defined as in Figs. 3a,b,c). A positive lag means that the dust variation occurs at larger latitudes than the ice variation. The correlations were computed in 256 bins covering the range from -36 to +36 degrees of latitude. The scale of CCF amplitudes (optical depth squared) are the same in both panels, so the degrees of correlation can be directly compared.

## 4. Discussion

### 4.1 MARCI daily water opacity maps

Figure 5 shows a sequence of averaged MARCI maps showing the water ice opacity in the north (top) and south (bottom) polar regions. Large dark areas delineate regions where no data was taken due to low or no sunlight.

Figure 6 shows MARCI images of the entire Martian globe displaying cloud conditions on particular Martian days in MY 29, across northern summer. They run from $L_s$=86.7-180. The last two images in the sequence show the build up of the north polar hood, the first six show the exposed north polar cap and the aphelion cloud belt (ACB) encompassing the equatorial regions. Individual water ice baroclinic or vortex storms can be seen moving north from the ACB in the $L_s$=120 and 132 images.

Figure 7 shows a snapshot of the north polar MARCI water ice observations for summer of Mars Year 29. The area this is taken from is shown in a red box in Figure 6. Each pixel in the image is 200 pixel average of along latitude 270-300E, which is beneath Gemini Lingula in the north polar cap.

Several observations which are relevant to this project can be made from this set of images in Figures 5-7. The presence of the polar hoods during winter is clear in this image set, as well as discrete cloud events in the north polar region which are generally absent from the south pole.

**Interannual repeatability**. Figure 5 shows the presence of the north polar hood and baroclinic cloud activity over several Martian Years that is the hallmark of the interannual repeatability we detected using temporal cross-correlation functions of water ice and dust. Typical lag ice events by 35-80 degrees of solar longitude in both poles. This is likely to be related to seasonal activity seen in Figure 5 and Figure 2 (for dust).

**Water ice event core and halos**. Both polar hoods show their greatest intensity on a belt exterior to the pole – this is clearest for the southern polar hood in the final image. There is some year to year variability, but similar patterns occur each year. This reinforces our finding that water ice events have smaller cores than dust events, and ice events in the north and south have roughly similar (10°) latitudinal extent.

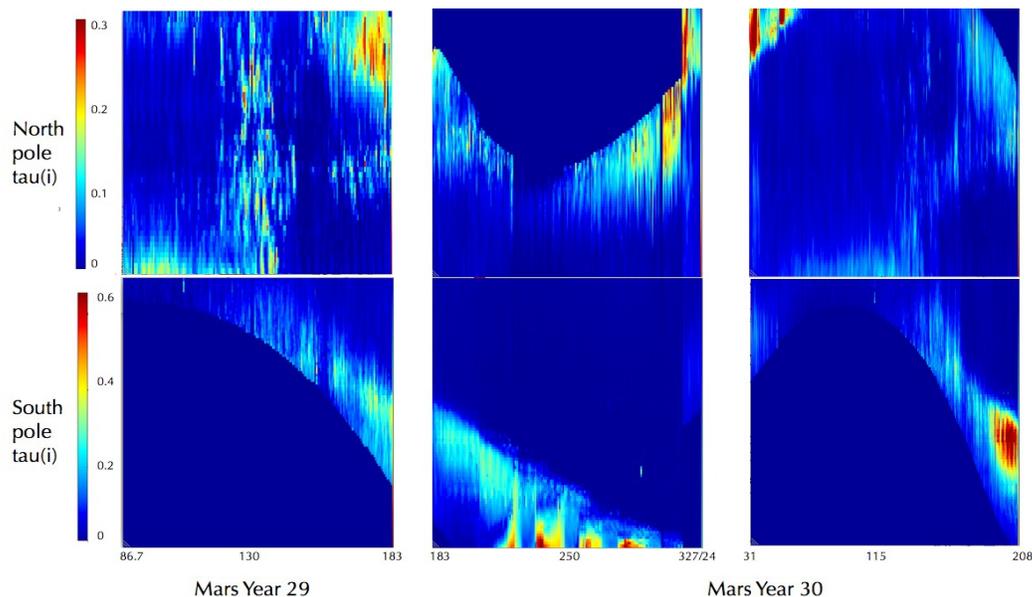

**Figure 5.** Sequence of averaged MARCI water ice maps: the top figure shows the north polar data in the top row and the south polar data in the second row. The areas of dark blue are over the polar caps during winter where no observations were possible. The dataset runs over 1.5 Martian years. The image run from 80-20N (top) and 20-80S (bottom).

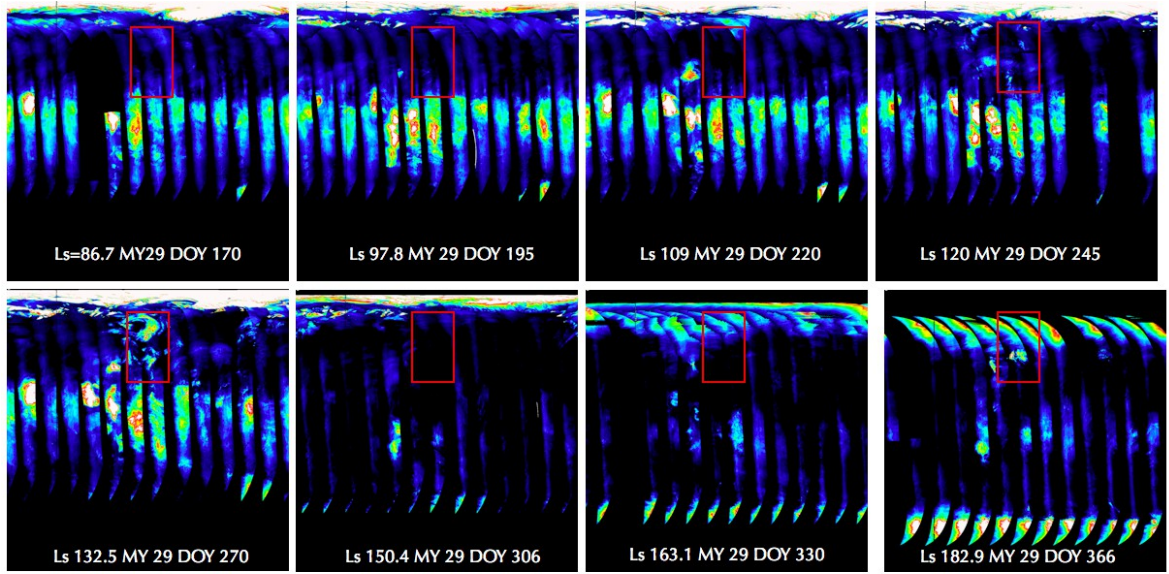

**Figure 6.** *Global MARCI maps for eight days throughout Mars Year 29, showing annual variations in water ice abundance. The red box shows the region focused on in Figure 6.*

### 4.2. Background Water ice opacity

It should be noted that the MARCI retrievals used in the correlation study reported here are only those corresponding to the CRISM EPF data. Nevertheless, there does appear to be a peak in water ice opacity in due to the polar hood that reaches similar opacities (our model suggests peaks of $\tau_i$ =1.5) in each pole that diminishes gradually during springtime. Background water ice is much lower than background dust opacity for both poles (consistently in 'background' measurements, $\tau_i < 0.1 < \tau_d$).

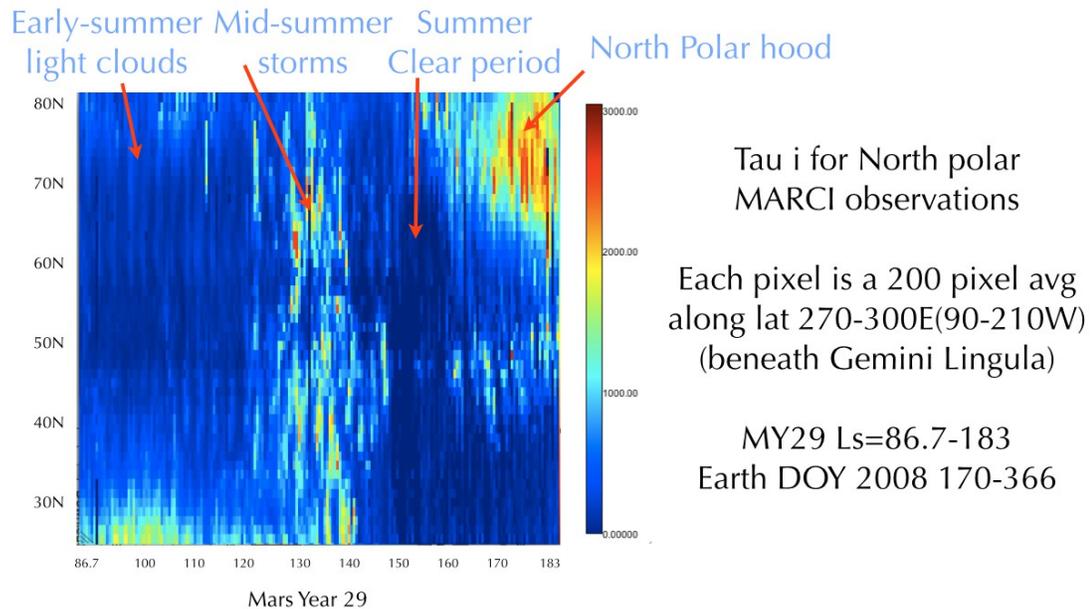

**Figure 7.** A snapshot of the north polar MARCI water ice observations for summer of Mars Year 29. The area this is taken from is shown in a red box in Figure 6.

### 4.3 Comparison with MCS polar hood datasets

Benson et al. (2010, 2011) reported maximum optical depths for the northern and southern polar hood of $\tau_i$ = 0.66. These values were derived from extrapolations of opacity profiles, where the lower 10km of the atmospheric column cannot be well constrained in the presence of even moderate amounts of atmospheric aerosols (e.g. (Kleinböhl et al., 2009)). Recent Global Climate Model results suggest that an appreciable amount of the atmospheric water ice column may be found in the bottom scale height during these seasons (R. Haberle, 2013, private communication). This fact and the various uncertainties associated scaling the MCS IR retrievals to visible wavelengths indicate that the MARCI optical depths reported here do not represent a fundamental mismatch: $0 < \tau_i < 1.5$ for the north pole and $0 < \tau_i < 2.0$ (often above $\tau_i$ =1) for the south pole (see Figure 2).

## 5. Conclusions

We have presented estimates of dust opacity from CRISM and ice cloud opacity from MARCI in the north (Figure 2) and south pole (Figure 3) for the first three Mars years of MRO operations (MY28-30), which included a large dust event at MY28/$L_s$=260-270 and a smaller dust event in MY29 at the same time period.

We have carried out a correlation analysis of the combined CRISM dust and MARCI cloud opacity datasets, which has lead to the following major results:

a.) **Flushing of southern dust storms**. In the southern polar region, dust storms have negative temporal autocorrelation over the range of lags between 70 and 110 degrees of solar longitude. This is related to the time it takes for dust storms to repeat.

b.) **Spatial and temporal anticorrelation of dust and ice in south polar region**. In the south polar region, dust and ice are temporally and spatially anti correlated. This is due to dust being raised in the spring, and water ice being present in the winter. In the north polar region, this relationship is reversed, however temporal correlation of northern dust and ice clouds is weak – 6 times weaker than the anticorrelation in the south polar region (Figure 4a).

c.) **Southern dust events are wider in latitudinal extent**. We have found that dust storms are more dispersed (they can cover the whole poles) than ice clouds (typically halos of less than 10°) in both poles. Southern dust events have a slightly wider extent than their northern cousins, and are typically more intense.

d.) **Dust activity lags cloud activity in both poles**. Using temporal cross-correlation functions of water ice and dust, we find that dust events temporally lag ice events by 35-80 degrees of solar longitude in both poles. This is likely to be related to seasonal activity.

e.) **Latitude cross-correlations between ice and dust in north**. Using the latitudinal cross correlation of dust and ice, we find that dust in the north lags ice in latitude. Physically, this corresponds to dust events closer to the pole by about 8° of latitude relative to ice events, and is likely related to katabatic dust events coming off the polar cap during late summer when early clouds of the north polar hood are present.

## 6. Acknowledgements

Our thanks to Scott Murchie (CRISM PI) and the CRISM science operations team at JHU APL for their dedication to the task of obtaining this unique dataset. We thank the MARCI team led by Mike Malin at MSSS. Our thanks to Mike Smith for providing his coork files for Martian $CO_2$ profiles. This work was funded under NASA Mars Data Analysis Program grants NNX11AN41G and NNX13AJ73G administered by Mitch Schulte.

All CRISM data used in this paper is publicly available at the Planetary Data System (PDS) Geosciences node. For derived dust opacity data, email the author at abrown@seti.org. All water ice opacity maps are available at the Mars Water Ice Wiki administered by MJW:
https://gemelli.spacescience.org/twiki/bin/view/MarsObservations/MarciObservations/WaterIceClouds